\documentclass[usenatbib]{mnras}

\usepackage{newtxtext,newtxmath}
\usepackage[T1]{fontenc}
\usepackage[utf8]{inputenc}
\usepackage{ae,aecompl}
\usepackage{hyperref}

\usepackage[usenames, dvipsnames]{color}
\usepackage{graphicx}	\usepackage{amsmath}	\usepackage{amssymb}	\usepackage{pifont}

\newcommand{\kpc}{\ensuremath{\, \mathrm{kpc}}}
\newcommand{\Mpc}{\ensuremath{\, \mathrm{ Mpc}}}

\newcommand{\Msun}{\ensuremath{\, \mathrm{M}_{\odot}}}

\newcommand{\Gyr}{\ensuremath{\, \mathrm{Gyr}}}

\newcommand{\ie}{\emph{i.e.}\, }
\newcommand{\eg}{\emph{e.g.}}

\renewcommand{\d}{\mathrm{d}}

\newcommand{\Mpri}{\ensuremath{\, M_\textrm{pri}}}
\newcommand{\Msec}{\ensuremath{\, M_\textrm{sec}}}
\newcommand{\zpri}{\ensuremath{\, z_\textrm{pri}}}
\newcommand{\taumerger}{\ensuremath{\, \tau_\mathrm{merger}}}
\newcommand{\SFRsec}{\ensuremath{\, SFR_\textrm{sec}}}
\newcommand{\SFRpri}{\ensuremath{\, SFR_\textrm{pri}}}

\newcommand{\hugo}[1]{#1}

\title[]{Real galaxy mergers from galaxy pair catalogs}

\author[H. Pfister et al.]{
Hugo Pfister,$^{1,2}$\thanks{Sophie and Tycho Brahe Fellow; hugo.pfister@nbi.ku.dk}
Massimo Dotti,$^{3,4}$
Clothilde Laigle,$^{5}$ \newauthor
Yohan Dubois$^{5}$
and Marta Volonteri$^{5}$\\
$^{1}$DARK, Niels Bohr Institute, University of Copenhagen, Denmark\\
$^{2}$University of Hong-Kong, China\\
$^{3}$Dipartimento di Fisica G. Occhialini, Universit$\grave{a}$ degli Studi di Milano--Bicocca, Piazza della Scienza 3, I-20126 Milano, Italy\\
$^{4}$INFN, Sezione Milano--Bicocca, Piazza della Scienza 3, I-20126 Milano, Italy\\
$^{5}$Institut d’Astrophysique de Paris, Sorbonne Universit\'e, CNRS, UMR 7095, 98 bis bd Arago, 75014 Paris, France\\
}

\date{Accepted XXX. Received YYY; in original form ZZZ}

\pubyear{2019}

\begin{document}

\label{firstpage}
\pagerange{\pageref{firstpage}--\pageref{lastpage}}
\maketitle

\begin{abstract}
Mergers of galaxies are extremely violent events shaping their evolution. Such events are thought to trigger starbursts and, possibly, black hole accretion. Nonetheless, it is still not clear how to know the fate of a galaxy pair from the data available at a given time, limiting our ability to constrain the exact role of mergers. In this paper we use the lightcone of the \textsc{Horizon-AGN} simulation, for which we know the fate of each pair, to test three selection processes aiming at identifying true merging pairs. We find that the simplest one (selecting objects within two thresholds on projected distance $d$ and redshift difference $\Delta z$) gives similar results than the most complex one (based on a neural network analyzing $d$, $\Delta z$, redshift of the primary, \hugo{masses/star formation rates/aspect ratio of both galaxies}). Our best thresholds are $d_\mathrm{th}\sim100\kpc$ and $\Delta z_\mathrm{th} \sim 10^{-3}$, in agreement with recent results.
\end{abstract}


\section{Introduction}

Galaxy interactions and mergers have been advocated as one of the principal actors in galaxy evolution. \cite{Toomre_77} proposed mergers as responsible for the fast morphological transformation of disc galaxies into spheroids or, in less dramatic cases, for the growth of massive classical bulges \citep{Hopkins_09c, Hopkins_09d}. Although this is still a debated result \citep{Fensch_17, Lofthouse_17}, gas-rich mergers have been proposed as triggers of intense bursts of star formation \citep{Barnes_91, Mihos_96, Cox_08, Calabro_19} resulting in luminous and ultra-luminous infrared galaxies \citep[][]{Sanders_88, Duc_97, Elbaz_03}, as well as triggers of high luminosity single and double active galactic nuclei \citep[AGNs,][]{DiMatteo_05, Capelo_15} possibly responsible for the quenching of star formation in the remnant \citep{Sijacki_06,  DiMatteo_08, Booth_09, Dubois_13}. Mergers can also lead to galaxy spin flip (from aligned to perpendicular) along filaments, therefore they bring diversity in the intrinsic alignment pattern \citep[\eg][]{Welker_14,Welker_19}. Finally, mergers of massive galaxies are the natural path to the formation of massive black hole pairs and binaries \citep[][]{Begelman_80, Tremmel_15, Tremmel_17, Tremmel_18, Bellovary_19, Pfister_17, Pfister_19a}. 
If the interaction with their complex environment leads the two black holes to \hugo{separation} of $10^{-3} (M_{\rm binary}/10^6 \Msun)^{0.75}$ pc\footnote{This estimate applies to close to equal mass circular binaries, for the discussion on the actual dependencies on the eccentricity and mass ratio see Eq.~(2) in \protect{\cite{Dotti_12}}.}, with $M_{\rm binary}$ the total mass of the black hole binary, it can further shrink and finally coalesce in less than an Hubble time, while emitting gravitational waves detectable by current and future observational campaigns \citep{Hobbs_10, AmaroSeoane_13, AmaroSeoane_17, Babak_16}.

For all these reasons, galaxy mergers and their consequences have been explored thoroughly from a theoretical point of view, both analyzing and post--processing the outcomes of coarse but large cosmological simulations \citep[\eg][]{Steinborg_16, Volonteri_16}, as well as higher resolution isolated mergers starting from idealized initial conditions \citep[\eg][]{Capelo_15}.

In order to confirm these results from an observational perspective, it is required to know which galaxies are going to merge. Two main methods have been used to obtain this information. The first relies on identifying perturbations in galaxy morphology due to mergers \citep{LeFevre_00, Conselice_03, Lotz_08, Goulding_18}. The second method, which we will study in more detail in this paper, is pair counting \citep{Zepf_89, LeFevre_00, Snyder_17, Snyder_18, Ventou_19, Duncan_19}: a pair is selected as ``merging'' if the relative projected distance ($d$) and redshift difference ($\Delta z$) of the two galaxies are smaller than given thresholds $d_\textrm{th}$ and $\Delta z_\textrm{th}$. Both methods have their advantages and drawbacks, in principle, the first one uses all the information in the images, but it requires very high resolution and therefore cannot be applied at high redshift. Pair counting uses ``less'' information and can be applied to higher redshift, but pairs with a large real 3D separations, which will not merge nor interact within a Hubble time, could be selected. The last point naturally raises the question of the optimal thresholds as well as the dependence of these thresholds with other parameters such as the masses, the mass ratio etc...

In this study we take full advantage of the results of the {\sc Horizon-AGN} cosmological simulation \citep{Dubois_14a} to build mock catalogues of observationally selected galaxy pairs and define the best technique to select merging pairs. In \S \ref{sec:BuildTheCatalog}, we detail how we construct this catalog and compare it with similar catalogs \citep{Snyder_17}; in \S \ref{sec:DetectionOfRealGalaxyMerger} we test three different algorithms to detect pairs and compare their efficiency; we finally  give our conclusions in \S \ref{sec:conclusions}.

\begin{figure}
\begin{center}
 \includegraphics[width=\columnwidth]{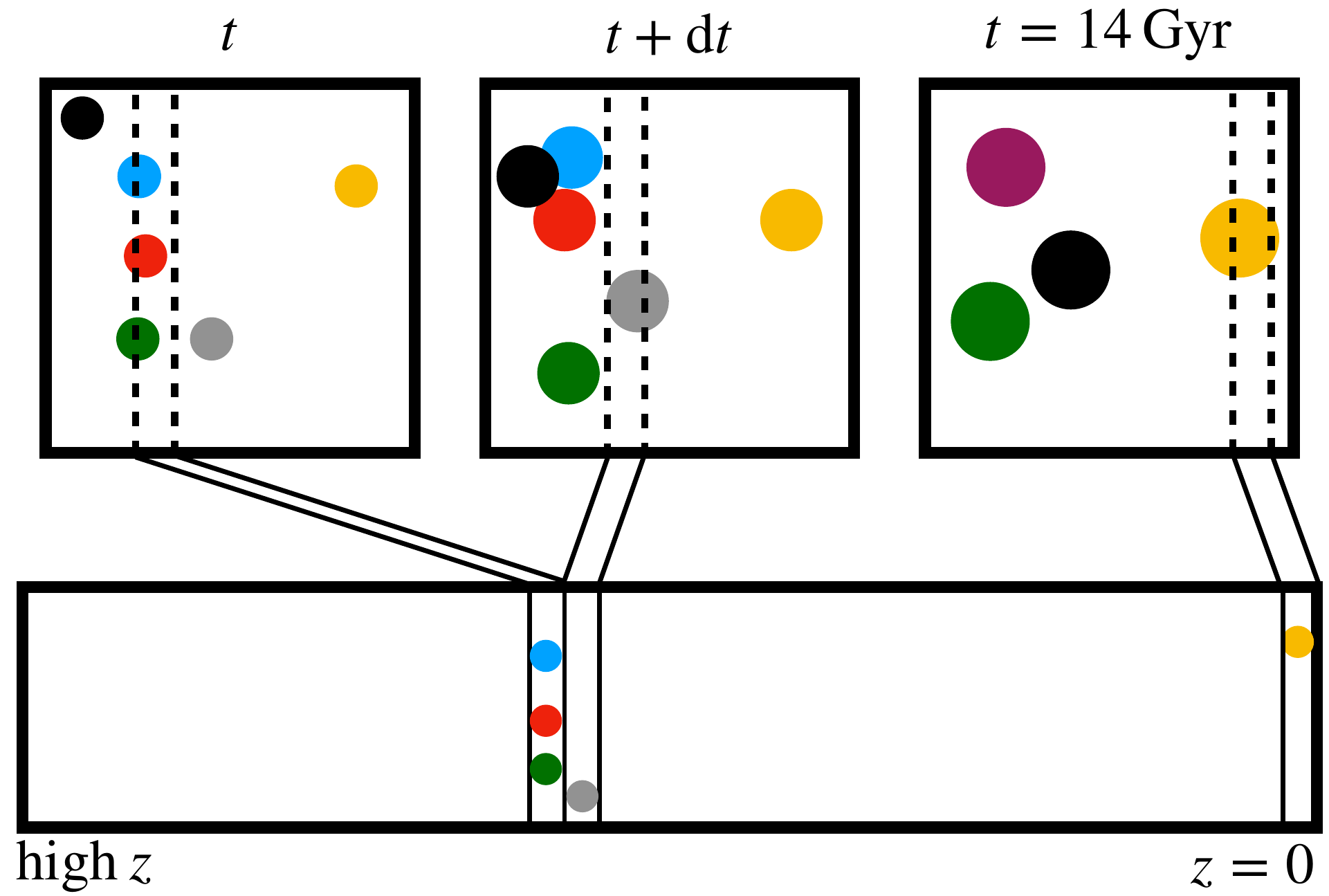} \\
 \rule{\columnwidth}{0.1cm}
  \includegraphics[width=\columnwidth]{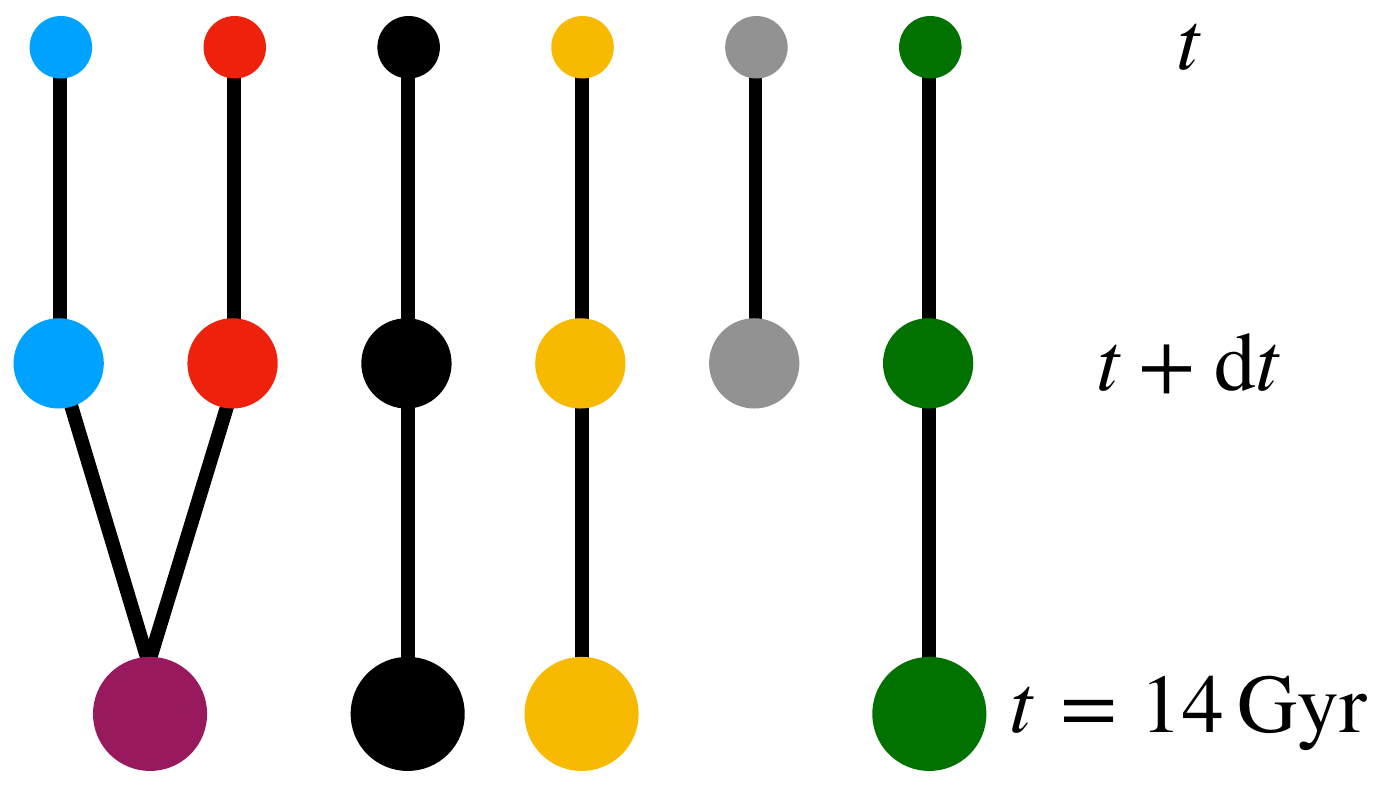} 
 \caption{\textbf{Top:} Sketch of the construction of a lightcone. Squares on the top line represent the simulated box, which is evolved in time. Slices of the box are stored at each timestep and are then stacked to form the lightcone (bottom line). Note that some galaxies are present at all time in the box, but not in the lightcone, and that some galaxies (the black one here) can be in the box without being in the lightcone. \textbf{Bottom:} Merger tree associated with the simulation sketched, some galaxies merge (blue-red pair), some remain isolated for a long time (yellow, black and green) and some ``dissolve'' as they have no child identified (grey).}
 \label{fig:lightcone}
 \end{center}
\end{figure}

\section{Build a numerical catalog}
\label{sec:BuildTheCatalog}

Our aim is to build a catalog of galaxy pairs, as an observer would do, but knowing, for a given pair, if it will merge or not. Here we detail how we build this catalog\footnote{Please contact the corresponding author if you are interested in obtaining the catalog.}. In \S \ref{sec:AvailableData}, we present the different data available we used; we then describe our method to build the catalog in \S \ref{sec:CompilingData}; finally, in \S \ref{sec:ValidationOfTheCatalog}, we compare this catalogs with results from \cite{Snyder_17} to verify its behavior.

\subsection{Available data}
\label{sec:AvailableData}
We use the data from the {\sc Horizon-AGN} simulation \citep{Dubois_14a}. This is one of the largest hydrodynamical cosmological simulation available, the box size is 140\Mpc\, at $z=0$, with 1 kpc resolution in the most refined regions and a dark matter particle mass of $8\times 10^7 \Msun$. It has been run with the adaptive mesh refinement code {\sc Ramses} \citep{Teyssier_02}, and contains state-of-the-art galaxy formation subgrid physics: cooling \citep{Sutherland_93}, background UV heating~\citep{Haardt_96}, star formation \citep{Rasera_06} resulting in a stellar particle mass of $2\times 10^6 \Msun$, feedback (stellar winds, type II and type Ia supernovae) and black hole formation, accretion and feedback~\citep{Dubois_12}. \textsc{Horizon-AGN} reproduces many properties of real galaxies \citep{Dubois_16, Volonteri_16, Kaviraj_17}, therefore we can use it to produce mock catalogs, from which we can derive realistic methods observers could use to interpret the data they collect.

\subsubsection{Galaxies in the lightcone}

Concentric shells centered on a fiducial observer located at the origin of the simulation box at $z=0$, and containing particles (dark matter, stars and black holes) as well as gas cells, have been extracted on the fly at each coarse time step of the simulation. This allows the creation of a lightcone \citep{Pichon_10,Gouin_19} as sketched in Fig.~\ref{fig:lightcone} (top). The opening angle is 2.25 deg from $z=0$ to $z=1$, corresponding to the angular size of the full simulation box at $z=1$.

A catalog of galaxies has been extracted from this lightcone~\citep{Laigle_17}  containing, in particular, the following information for each galaxy:
\begin{itemize}
\item stellar mass $M$;
\item star formation rate $SFR$;
\item \hugo{aspect ratio $\gamma$ as seen in the lightcone, defined as the ratio between the semi-minor and semi-major axis;}
\item location on the sky with right ascension and declination;
\item observed redshift, $z$, corresponding to the redshift an observer would measure from a spectroscopic dataset.
\end{itemize}

\subsubsection{Galaxies in the box}
Galaxies in the box have been identified with \textsc{AdaptaHOP} \citep{Aubert_04}. The algorithm detects gravitationally bound structures containing at least 50 stellar particles, therefore having a minimum mass of $10^8$ \Msun. Using again the sketch in Fig.~\ref{fig:lightcone} (bottom), the blue, red, green, grey and yellow dots (galaxies) are now identified both in the box and in the lightcone. However, initially, galaxies in the lightcone and in snapshots are not matched. This matching is important because, for galaxies in the lightcone, similarly to galaxies in the sky, we have only an image at one particular time. Galaxies in the box are instead consistently evolved from $z = 100$ down to $z = 0$, therefore we know their history, \eg we know how they move or how their mass evolves (as represented by the enhancement of the size of the dots). Each galaxy that can be observed in the lightcone has been associated to the same galaxy in the box~\citep{Laigle_17}, connecting the ``observational view'' to how a galaxy actually evolves over cosmic time.

\subsubsection{Merger tree}
The merger tree of galaxies in the box has been produced with \textsc{TreeMaker} \citep{Tweed_09}. Galaxies containing particles with the same ID at different times are matched to form the history of each galaxy. With this we can follow galaxies from their birth down to $z=0$. This is sketched in Fig.~\ref{fig:lightcone}  (bottom) where galaxies are followed with time, some merging (blue-red pair), some remaining isolated (black, yellow and green dots) and some ``dissolving'' (grey dot, see \S \ref{sec:CompilingData}).

\subsection{Combining data}
\label{sec:CompilingData}
With all the data presented in the previous Sections, for a pair of galaxies in the lightcone, we have a pair of galaxies in the box, which we can follow down to $z=0$ with the merger tree to see if they merge, or not, and how long it takes if it is actually the case. We select all the pairs (in the lightcone) fulfilling the following criteria:
\begin{itemize}
\item Both galaxies must be observed at redshift $0.05<z<1$, this is the redshift range from which the lightcone of {\sc Horizon-AGN} with angular opening $2.25\deg$ has been produced;
\item The mass of the most massive galaxy has to be larger than $10^9 \Msun$, so that it is defined with at least 500 stellar particles, and lower that $10^{11}\Msun$, so that there are more than 10 of those galaxies in the catalog. We also impose a stellar mass ratio between galaxies of $0.1<q<1$;
\item The projected distance between the two galaxies, $d$, measured with the angular distance assuming that the redshift is the one of the primary galaxy, has to be lower than $5 \Mpc$. Similarly, the redshift difference between the two galaxies, $\Delta z$, has to be lower than $0.05$. These criteria are intentionally extremely loose to ensure that most merging pairs are included. Consistently with the simulation, we assume a $\Lambda$CDM cosmology with \emph{WMAP7} parameters \citep{Komatsu_11}.

\end{itemize}

We end up with $\sim 9\times 10^8$ pairs of galaxies, for which we know the following observational quantities: the mass of the primary and mass ratio, $M_\mathrm{pri}$ and $q$; the redshift difference, $\Delta z$; the projected distance, $d$; their SFR, \SFRpri and \SFRsec; \hugo{and their aspect ratios, $\gamma_\mathrm{pri}$ and $\gamma_\mathrm{sec}$}. We also know the associated pair in the snapshots, which we can follow in the merger tree. We use the sketch in Fig.~\ref{fig:lightcone} (bottom) to list the possible cases:
\begin{enumerate}
\item The two galaxies live at the same time, \ie they are in the same snapshot. We can then follow their history in the merger tree and see if the two galaxies merge, and how long it takes (\taumerger). For instance, the blue-red pair merges, while the blue-green pair has not merged by $z=0$ ($\taumerger = \infty$).
\item  The two galaxies do not live at the same time, \ie they are not in the same snapshot. We then follow the history of the galaxy with higher redshift until the two galaxies are at the same snapshot, and then apply case (i). For instance we would trace the blue galaxy in the blue-grey pair until time $t+\d t$ and then apply case (i). In this particular example, an additional feature happens: the grey galaxy ``dissolves'', this can happen if the galaxy \hugo{loses} enough stars, or is so perturbed, that it is not recognized by \textsc{AdaptaHOP} in one snapshot. As it is difficult to differentiate between a numerical and a physical disruption, pairs in which a galaxy ``dissolves'' are discarded (this represents $\sim 3\times 10^7$ pairs, a small fraction of the total).
\end{enumerate}

\subsection{Validation of the catalog}
\label{sec:ValidationOfTheCatalog}
To confirm that our catalog is coherent with previous studies, we perform a similar analysis as done for Fig.~2 in \cite{Snyder_17}: at a given redshift $\zpri$ for the primary, we estimate how many pairs fulfill the criterion observers use to define a merger, \emph{i.e.}  $\Delta z < 0.02(1+\zpri) = \Delta z_\textrm{th} $ and $d<75 \kpc=d_\textrm{th}$. Given this selection process, we can count how many selected pairs actually merge (true positive, $TP$) and how many selected pairs actually do not merge by $z=0$ (false positive, $FP$). From this, we compute the purity $P$ in $[0,1]$, corresponding  to the fraction of selected pairs that actually merge:
\begin{eqnarray}
P=\frac{TP}{TP+FP} \, .
\end{eqnarray}

\begin{figure}
\begin{center}
\includegraphics[width=\columnwidth]{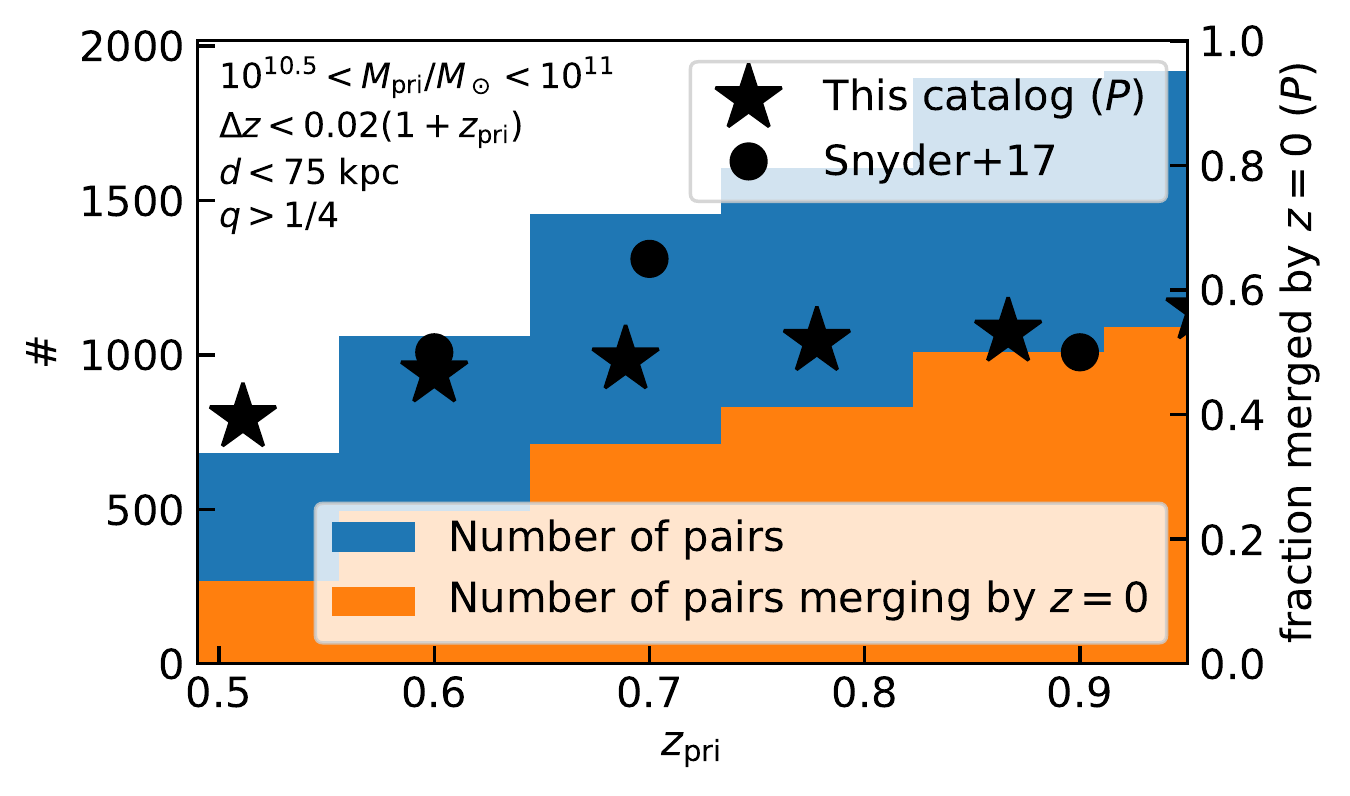}
\caption{\textbf{Left axis, histograms}: Total number of seletected pairs (blue) and number of pairs actually merging by $z=0$ (orange). We show in the top left corner the criterion to define a pair. \textbf{Right axis, markers}: purity of the selection in our catalog and in \protect\cite{Snyder_17}.}
\label{fig:FracMergerByzEqualZero}
\end{center}
\end{figure}

We show our results in Fig.~\ref{fig:FracMergerByzEqualZero}. Similarly to \cite{Snyder_17}, we find that, for $\zpri<1$, about 50\% of the pairs selected with $\Delta z < 0.02(1+\zpri)= \Delta z_\textrm{th} $ and $d<75 \kpc=d_\textrm{th}$ will not have merged by $z=0$. This confirms the robustness of this results, the goodness of our catalog, and at the same time, shows that there is room from improvement in detecting true mergers from galaxy pairs \citep{Cibinel_15,Snyder_18}.

\section{Detection of real galaxy mergers}
\label{sec:DetectionOfRealGalaxyMerger}

In this Section, we test three algorithms to detect real galaxy mergers from the available properties of each pairs in the catalog. We first show in \S \ref{sec:TwoExtremeCases} what are the main problems that must be overcome to build a faithful catalog without \hugo{losing} too many real pairs; we then discuss the metric we will use to judge the quality of the algorithm in \S \ref{sec:Goodness}; and we finally detail the algorithms in \S \ref{sec:OptimalParametersForDetectingPairs} and \S \ref{sec:NN}.

From now on, we will not consider the ``number of pairs which have merged by $z=0$'', since it is is not representative of the instantaneous merger rate, we consider instead the ``number of pairs which merge within 3 Gyr'', meaning that $\taumerger~<~3 \Gyr~=~\tau_{\rm merger, max}$. Note that for pairs with $z<0.25$, the time left before $z=0$ is less than 3 Gyr, in that case we consider indeed ``pairs which have merged by $z=0$''. The value of $\tau_{\rm merger, max}=3 \Gyr$ has been chosen because it is in agreement with typical merger timescales obtained in numerical simulations \citep{Capelo_15}, but we stress that its exact value is rather arbitrary, and partially affects the results as shown in Appendix \ref{sec:ChangingOurDefitionOfMerger}.

\subsection{A difficult exercise}
\label{sec:TwoExtremeCases}

Of the $9\times 10^8$ pairs, only  $\sim 10^5$, \ie only $\sim 0.01\%$, merge: the problem of detecting merging pairs is extremely unbalanced. This fraction depends on the particular parameters we used to select pairs (we do not expect to have mergers for $d\sim5\Mpc$...), but it is expected to be always low, as most pairs in the sky do not merge.

In addition, the problem is also extremely degenerate. For instance, we show in Table~\ref{table:ExtremePairs} the details and fate of four specific pairs:
\begin{itemize}
\item Pairs 1 and 2 consist in two pairs, with similar properties in terms of projected distance and redshift difference, but, given our definition of ``merger'', one of them merges and the other does not;
\item Pairs 3 and 4 consist in two pairs with, in both cases, two galaxies very close in redshift space ($\Delta z \lesssim 10^{-3}$) but, in one case, the two galaxies are far given the projected threshold usually used (projected distance $d$ is 349 \kpc) and, in the other case, they are close ($d$ is 37 \kpc); nonetheless, the distant pair merges whereas the close one does not.
\end{itemize}

\begin{figure}
\begin{center}
\includegraphics[width=\columnwidth]{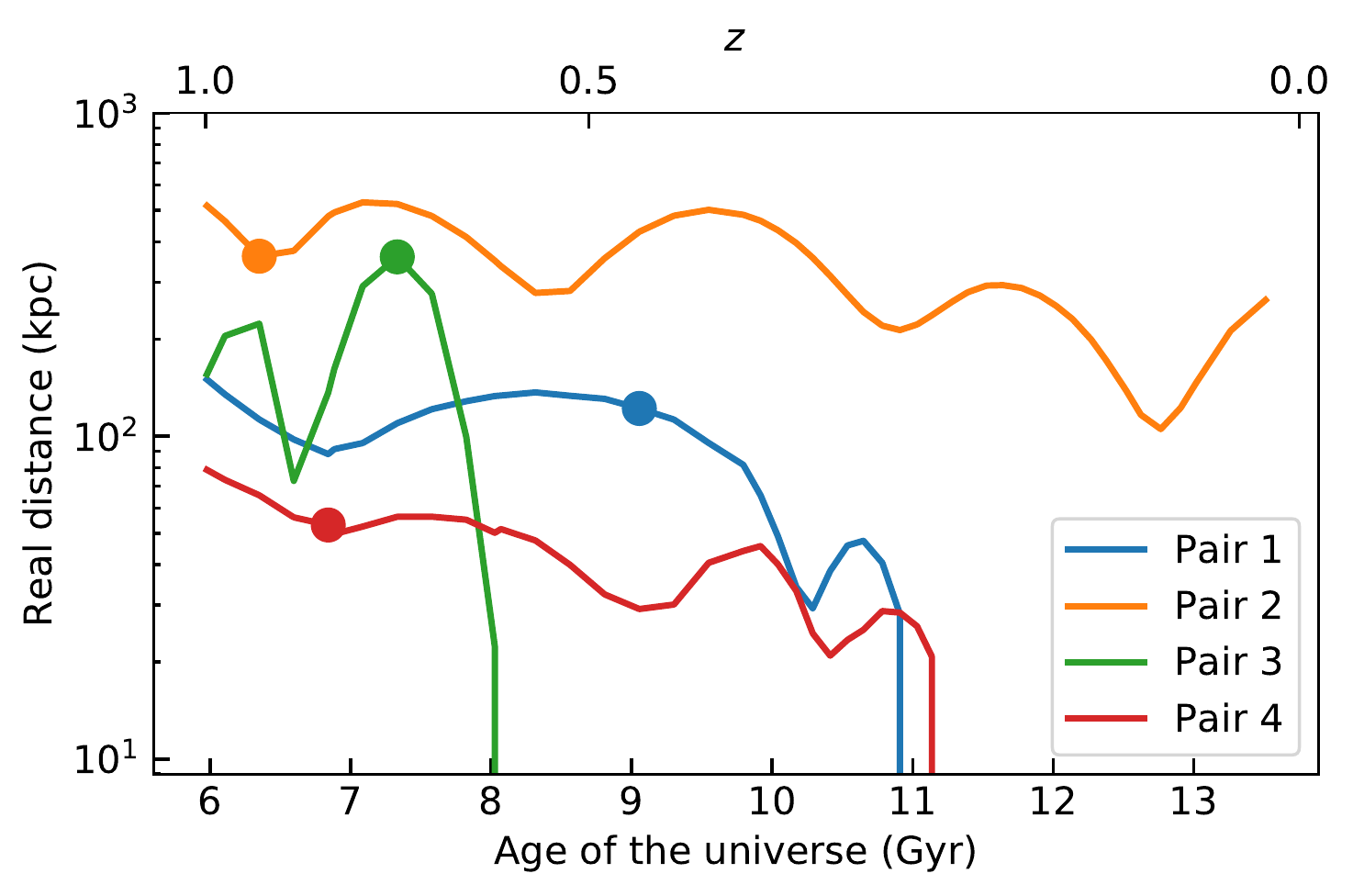}
\caption{3D distance as a function of time between the two galaxies of the four pairs described in Table~\ref{table:ExtremePairs}. Markers indicate the time at which the pair is seen in the lightcone.}
\label{fig:distances}
\end{center}
\end{figure}

We show in Fig.~\ref{fig:distances} the 3D distance (solid line), as measured in the simulation, between the two galaxies in the four pairs described in Table~\ref{table:ExtremePairs}, as a function of time. We also indicate at which time the pair is ``seen'' in the lightcone (marker). Pairs 1 and 2 have the same observed $d$ and $\Delta z$ but have in reality different orbital parameters. Galaxies in Pair 1 are separated by indeed $\sim100 \kpc$ and a relative speed of 100 km/s, but galaxies in Pair 2 are in fact separated by 400 kpc and a relative speed of 800 km/s. These different orbital parameters lead to a different fate, and a different merger timescale. Pair 3 is seen at the apocententer of a very eccentric orbit resulting in a fast merger, while Pair 4 is more circular, explaining why the merger takes a longer time.

From these simple examples, it is clear that using only thresholds on projected distance and redshift difference cannot be 100\% accurate, other quantities such as the masses (see \S\ref{sec:DependanceInMpri}), the shapes, the colors etc... or relations between all these quantities (see \S \ref{sec:NN}) should be used \citep[see also][]{Snyder_17,Goulding_18,Snyder_18}.

\begin{table}
\begin{center}
\begin{tabular}{ccccccc}
\hline
Pair ID & \zpri & \Mpri & \Msec &  $d$ & $\Delta z$ & \taumerger \\
&& $10^{10}$ \Msun & $10^{10}$ \Msun & \kpc & $10^{-4}$ & \Gyr\\
\hline
\hline
1 & 0.4578 & 2.307 & 2.05& 123&  0.4  & 1.977 \\
2 & 0.8996 & 1.293 & 0.485 & 125&  0.4  & $\infty$ \\
3 & 0.711 & 3.588 & 0.412 & 349 &  0.4   & 0.738 \\
4 & 0.7542 & 1.554 & 0.503 & 37&  2.0  & 6.427 \\
\hline
\end{tabular}
\caption{Four pairs, with similar properties but different behaviors.}
\label{table:ExtremePairs}
\end{center}
\end{table}

\begin{figure}
\begin{center}

\includegraphics[width=\columnwidth]{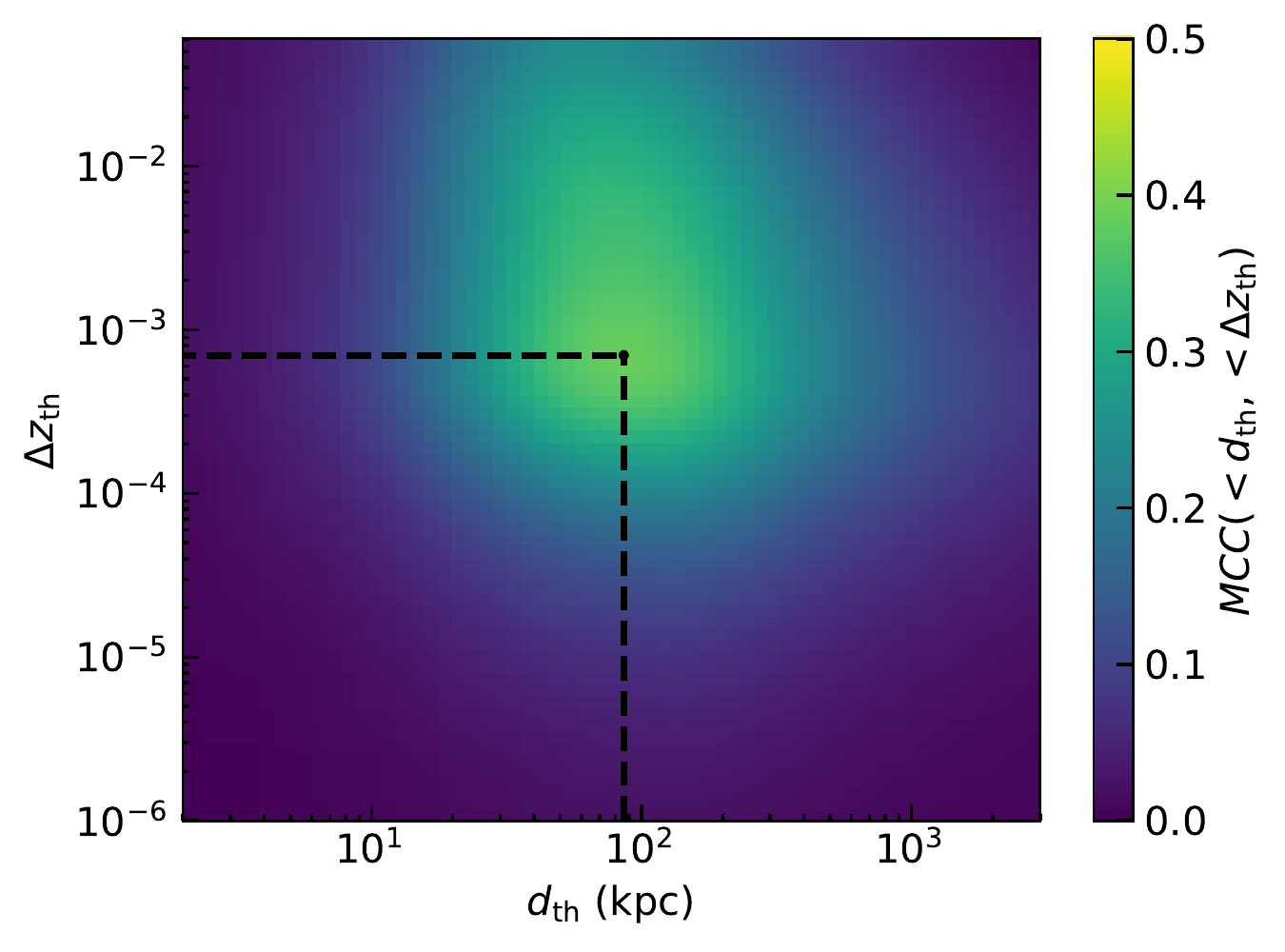}
\caption{Estimate of the $MCC$ as a function of the thresholds on redshift difference $\Delta z_\mathrm{th}$ and projected distance $d_\mathrm{th}$.}

\label{fig:PCMCCF1}
\end{center}
\end{figure}

\subsection{Goodness of the detection method}
\label{sec:Goodness}

To compare two selection methods, and judge which one is the best, we need a metric. Purity (see \S\ref{sec:ValidationOfTheCatalog}) only is not a good metric, as a very restrictive threshold (very small $d_\textrm{th}$ and $\Delta z_\textrm{th}$) would result in a purity of 100\%, but would miss many true mergers. This is why we also consider the completeness $C$ in $[0,1]$ corresponding to \hugo{the fraction of true mergers selected}:
\begin{eqnarray}
C=\frac{TP}{TP+FN} \, ,
\end{eqnarray}
where $FN$ (false negative) corresponds to the number of true mergers not selected.

Clearly, purity and completeness vary in opposite directions: if the thresholds are very restrictive, as we already said, purity will be high, but completeness will be low, and vice versa. For this reason, we need a combination of $P$ and $C$ or, similarly, of $FP$, $FN$, $TP$ and $TN$, where $TN$ (true negative) corresponds to the number of non mergers non selected. We use the \emph{Matthews correlation coefficient}, $MCC$ \citep{MCC}, defined as:
\begin{eqnarray}
MCC  &=& \frac{TP\times TN - FP\times FN}{\sqrt{(TP+FP)(TP+FN)(TN+FP)(TN+FN)}} \nonumber \, .\end{eqnarray}
The $MCC$ is in $[-1,1]$, 1 meaning that the algorithm gives perfect predictions, 0 meaning that it is random and -1 meaning it is always wrong. 

\subsection{Using simple thresholds on $d_\textrm{th}$ and $\Delta z_\textrm{th}$}
\label{sec:OptimalParametersForDetectingPairs}

\subsubsection{Starting point}
\label{sec:DependanceInDandZ}

We begin with a simple detection method: a pair is selected and (observationally) defined as merging if its projected distance and redshift difference are lower than the thresholds $\Delta z_\textrm{th}$ and $d_\textrm{th}$. As discussed in \S \ref{sec:TwoExtremeCases} this method cannot be 100\% accurate but it still is a reasonable (and frequently used) starting point.

In this Section, we search for the best $\Delta z_\textrm{th}$ and $d_\textrm{th}$ that optimize the $MCC$. For this purpose, we vary the two parameters and compute the $MCC$.

In Fig.~\ref{fig:PCMCCF1}, we show the $MCC$ given the thresholds used. We marked with a dashed-black line when this metric is at maximum. We find $\Delta z_\textrm{th}=7\times10^{-4}\sim  10^{-3}$ and $d_\textrm{th} = 86 \kpc \sim 100 \kpc$. The value of $100 \kpc$ is similar to the threshold used by observers, and it is of the order of magnitude expected: galaxies separated by 10 kpc are very likely to undergo a merging process \citep[the typical scale length of the disc of the Milky-Way is 3.5 kpc,][]{BT_87}; and 1000 kpc would correspond to very distant, probably non interacting, pairs. The value of $10^{-3}$ for the redshift difference, hardly achievable with current photometric redshift, is however about one order of magnitude lower than the threshold usually chosen, typically $10^{-2}(1+\zpri)$. Note that \cite{Pasquet_19} suggested a method allowing to reach $10^{-3}$ uncertainty on photometric redshift measurement, which is encouraging for the future surveys.

The best thresholds give $P=0.36$ and $C=0.41$. This again confirms our first guess of \S\ref{sec:TwoExtremeCases}: using only $d$ and $\Delta z$ is too degenerate to properly distinguish between mergers and non mergers. While the value of $P=0.36$ might seems extremely low, we recall here that the catalog is extremely unbalanced and degenerate, with only 0.01\% of pairs actually merging, therefore this selection method is actually three orders of magnitude better than random selection. The maximal $MCC=0.38$ is surprisingly good given the simplicity of the method. For example, \cite{Snyder_18}, using Random Forest on 10 parameters on pairs from the \textsc{Illustris} simulation \citep{Vogelsberger_14}, find a $MCC$ of about 0.4, with the difference that they considered pairs up to $z \sim 9$.

It is interesting to note that, at the time we were writing this paper, \cite{Ventou_19} performed a similar independent analysis to determine the optimal thresholds to detect merging pairs. They use a different simulation \citep[\textsc{Illustris}, ][]{Vogelsberger_14}, a different redshift range (up to $z=5$ but with lower redshift resolution as they use 6 snapshots), and a different metric to chose their threshold (completeness of 30\%), nonetheless, they found similar values, with pairs selected as merging if $d<50\kpc$ and $\Delta z < 10^{-3}$ or $50\kpc < d < 100 \kpc$ and $d<3\times 10^{-4}$. This supports our and their findings.

\begin{figure}
\begin{center}
\includegraphics[width=\columnwidth]{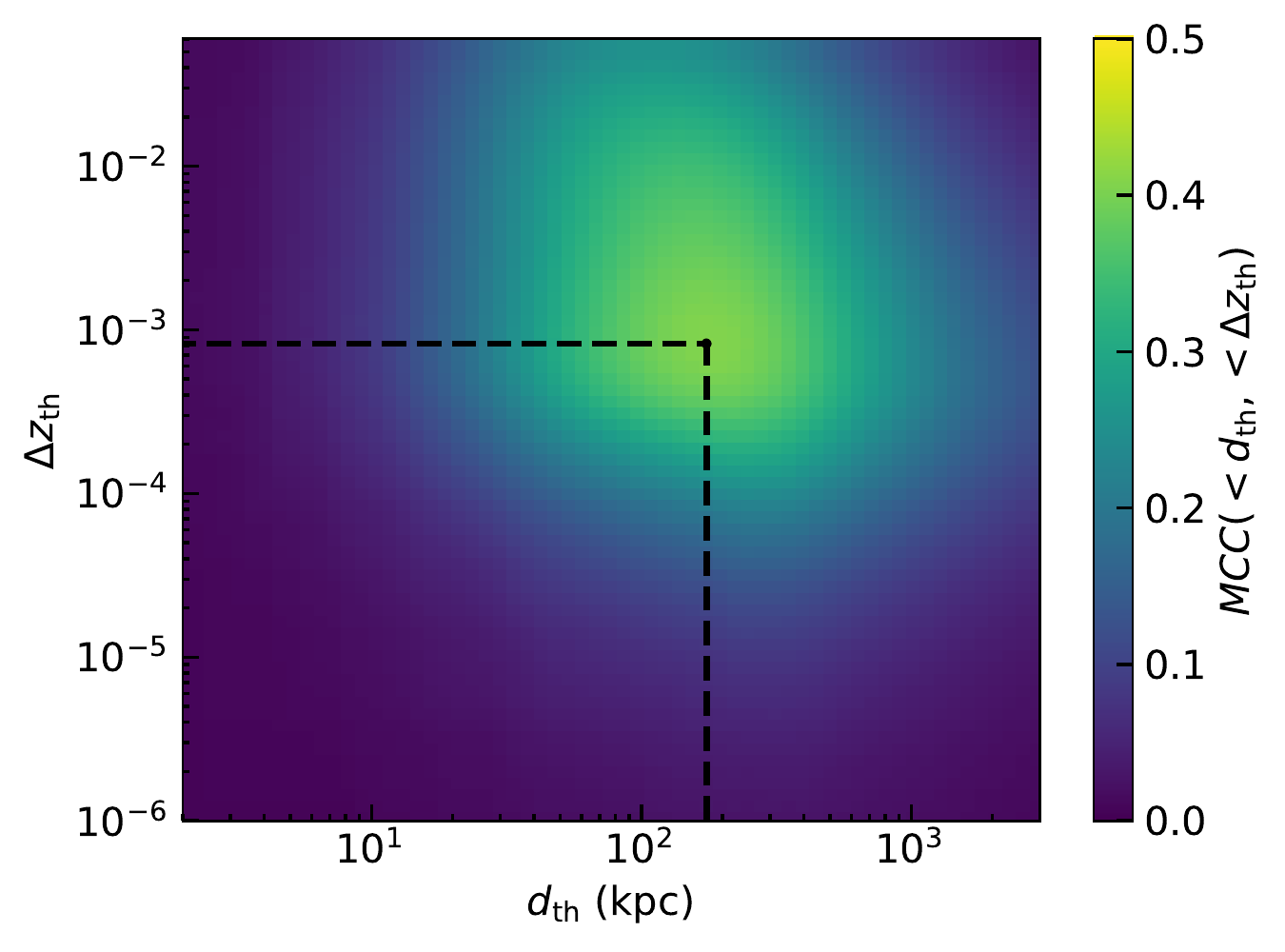}
\hfill
\includegraphics[width=\columnwidth]{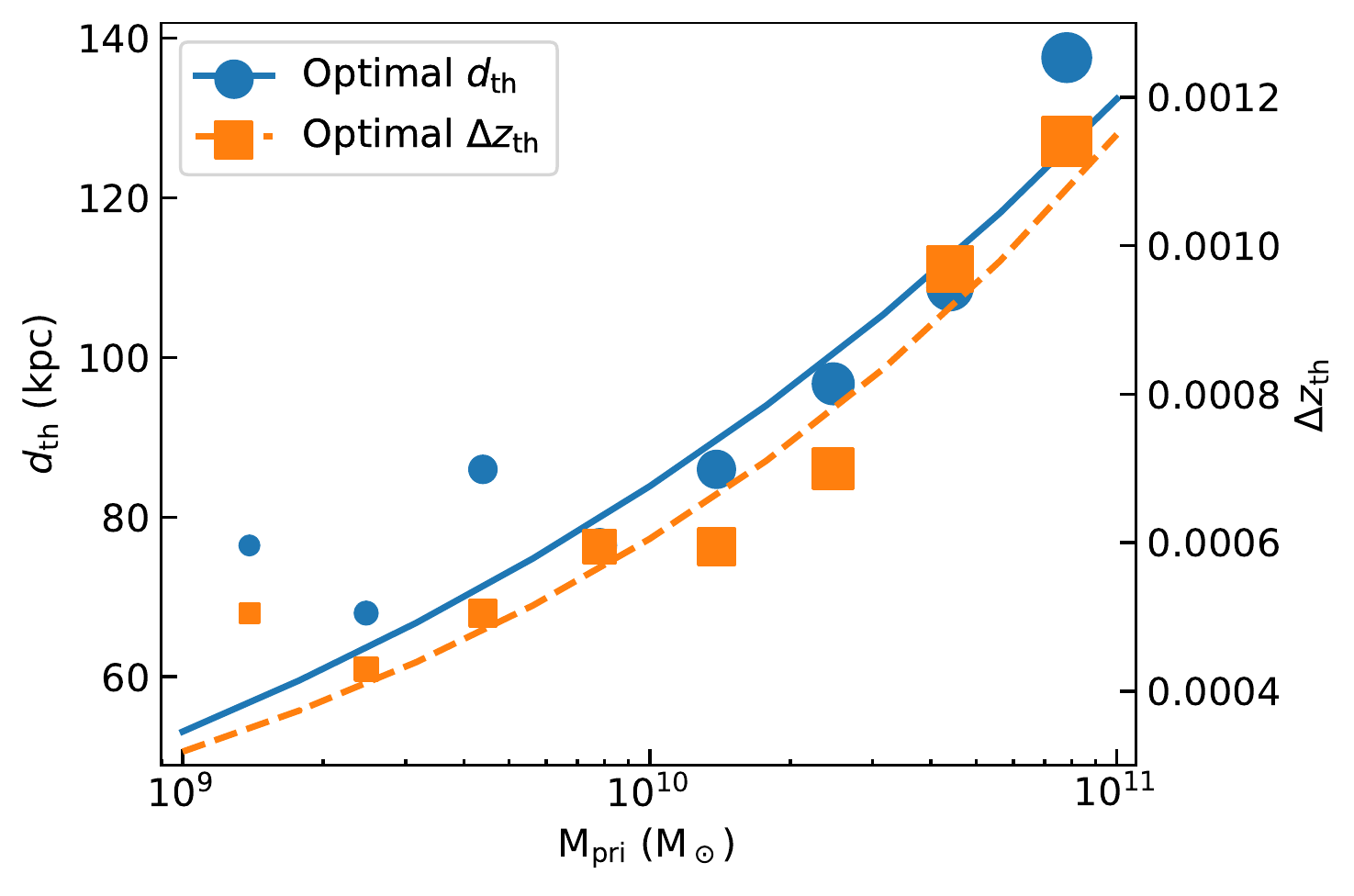}\\
\caption{\textbf{Top:} Estimate of the $MCC$ as a function of the thresholds on redshift difference $\Delta z_\mathrm{th}$ and projected distance $d_\mathrm{th}$. In this example, $\Mpri$ is in between  $10^{10}$ and $10^{10.25}$ \Msun. \textbf{Bottom:} Evolution of the optimal $d_\mathrm{th}$ (blue dots) and $\Delta z_\mathrm{th}$ (orange squares) for $M_\textrm{pri}$ within different mass intervals equally spaced by 0.25 dex, as well as their fits reported in Eq.~\eqref{eq:dthOfMpri} and Eq.~\eqref{eq:dzthOfMpri}. The larger the dots the larger the $MCC$ (min: 0.28, max: 0.54).}
\label{fig:PCofdr}
\end{center}
\end{figure} 
 
\subsubsection{Including the dependence on $M_\mathrm{pri}$}
\label{sec:DependanceInMpri}

More massive galaxies are expected to merge more frequently \hugo{than} low mass galaxies \citep{Fakhouri_10}. This is why we expect the thresholds $\Delta z_\textrm{th}$ and $d_\textrm{th}$ to depend on the masses of galaxies.

We explore this with our second selection method: a pair is selected and (observationally) defined as merging if its projected distance and redshift difference are lower than the \Mpri\, dependent thresholds $\Delta z_\textrm{th}(\Mpri)$ and $d_\textrm{th}(\Mpri)$.

With this idea in mind, we split the catalog in sub-catalogs in which \Mpri\, is in $[M_\mathrm{min}, M_\mathrm{max}]$, where $M_\textrm{min}$ ($M_\textrm{max}$) varies in equal logarithmic bins (0.25 dex) in between $10^9$ and $10^{11}$ \Msun. We then vary $d_\mathrm{th}$ and $\Delta z_\mathrm{th}$, and compute the $MCC$, as shown on the example in Fig.~\ref{fig:PCofdr} (top). In this example, for a primary mass within $10^{10}$ and $10^{10.25}$ \Msun, the $MCC$ peaks at 0.46 for $d_\mathrm {th}= 86 \kpc$ and $\Delta z_\mathrm {th}= 6\times 10^{-4} $: the classification is much better than using only $d_\mathrm{th}$ and $\Delta z_\mathrm{th}$. However, both the $MCC$ and the thresholds depend on \Mpri. We show this dependence in Fig.~\ref{fig:PCofdr} as well as a simple power law fit for the evolution of the thresholds:
\begin{eqnarray}
d_\mathrm{th} &=& 84 \kpc \left( \frac{\Mpri}{10^{10}\Msun} \right)^{0.20} \label{eq:dthOfMpri} \\
\Delta z_\mathrm{th} &=& 6\times 10^{-4} \left( \frac{\Mpri}{10^{10}\Msun} \right)^{0.28} \, . \label{eq:dzthOfMpri}
\end{eqnarray}

Both thresholds logarithmically scale as $\sim 1/3 \times \log(\Mpri)$. This is expected, indeed, using Eq.~(12) from \cite{Taffoni_03}, the dynamical friction timescale \citep{Chandrasekhar_43, BT_87}, which we use as a proxy for $\taumerger$, scales as:
\begin{eqnarray}
\taumerger = \frac{r^2 v_{c,\mathrm{pri}}}{\Msec}\, ,
\end{eqnarray}
where $v_{c,\mathrm{pri}}$ is the circular velocity at the virial radius of the primary and $r$ the real 3D distance between the galaxies. Considering constant density $\tilde{\rho}$ ($\Mpri \sim \tilde{\rho} R^3_\mathrm{pri}$) and virialized galaxies ($v_{c,\mathrm{pri}} \propto \Mpri^{1/2} R^{-1/2}_\mathrm{pri}$ ) immediately leads to $\taumerger \propto \Mpri^{-2/3}q^{-1}r^2$. In conclusions, for similar mass ratios ($q$ is between 0.1 and 1 in our catalog), and for a fixed dynamical friction timescale ($\taumerger < \tau_{\rm merger, max} =3 \Gyr$), we have $r\propto \Mpri^{1/3}$.

The $MCC$ is higher ($\sim 0.6$) for massive galaxies than for low mass galaxies ($\sim 0.3$): it is easier to detect real mergers of massive pairs. However, overall, if we chose $\Delta z_\mathrm{th}$ and $d_\mathrm{th}$ given by Eqs.~\eqref{eq:dthOfMpri} and~\eqref{eq:dzthOfMpri}, we find $P=0.43$, $C=0.36$ and $MCC=0.40$. \hugo{Note that we have optimistically assumed that $M_\mathrm{pri}$ is perfectly known, which is not true for real catalogs \citep[uncertainty on mass is typically 0.3~dex, ][]{Davidzon_17}}. In conclusions, including \Mpri\, results in a minor improvement of the classification compared with selection using only $\Delta z $ and $d$.

\subsection{Using a neural network}
\label{sec:NN}

We have shown in \S \ref{sec:DependanceInMpri} that using additional information than the projected distance or redshift difference can marginally improve the quality of the detection method. Similarly, non-linear relations between all properties of each pair could improve the quality of the detection method. 

We explore this with our third selection method: we build a simple neural network with {\sc keras} \citep{Keras}, which we train so that it detects merging pairs from the properties available in our catalog. Below we describe the main features of the network and the parameters used to ensure the reproducibility of our test. The architecture of the network is \hugo{somewhat} similar to the one from \cite{Marchetti_17}, with:
\begin{itemize}
\item \hugo{An input layer with the 9 parameters of each pairs ($d_\mathrm{th}$, $\Delta z_\mathrm{th}$, \Mpri, $q$, $z_\mathrm{pri}$, $SFR_\mathrm{pri}$, $SFR_\mathrm{sec}$, $\gamma_\mathrm{pri}$, $\gamma_\mathrm{sec}$)};
\item a first hidden layer in which 5 neurons, \ie 5 linear combinations of the \hugo{9 parameters resulting in 45 weights and 5 bias}. In order to introduce non-linearities, the results of these linear combinations are passed to an activation function for which we chose a hyperbolic tangent;\footnote{We tested a sigmoid activation function as well and found that the neural network behaved best with the hyperbolic tangent.} 
\item A second hidden layer, again with 5 neurons\footnote{In both hidden layers, we also tried with 10, 20, 50 and 100 neurons, which resulted in no significant improvement. Above 20 neurons the results actually becomes worse due to overfitting. For these reasons we finally opted for 5 neurons.}, \ie 5 linear combinations of the 5 outputs of the previous layer resulting in 30 new free parameters. Again, the results are passed to a hyperbolic tangent;
\item An output layer with one neuron, \ie a linear combination of the 5 outputs of the previous layer resulting in 6 new free parameters. In order to obtain a number that could be interpreted as a merging probability, the activation function chosen here is a sigmoid returning a real number $f$ in [0,1], where objects that the neural network considers secure non-mergers correspond to 0 while secure mergers are labeled 1.
\end{itemize}

The first step is the training of the network. For this task we used a sub-catalog (referred as \textit{training set}, with 1\% of the catalog, \ie $10^7$ pairs). The training proceeds running through this catalog multiple times (epochs), and evolving the \hugo{86} parameters of the 11 linear combinations computed in the hidden layers. As "loss function" (the metric used to evaluated how well each set of coefficients performs) we use the binary cross-entropy:
\begin{eqnarray}
L(y,\hat{y}) = -\frac{1}{N}\sum_{i=0}^N\left( y_i\log(\hat{y}_i) + (1-y_i)\log(1-\hat{y}_i) \right) \, ,
\end{eqnarray}
where $y$ corresponds to the real labels of the $N=10^7$ pairs of the \textit{training set}, $\hat{y}$ to the predicted label by the network, and index $i$ refer to a given pair. If a pair is merging (non merging) then its true label is 1 (0), if the predicted label is 1 (0) then $L$ will be null and if it is 0 (1) then the loss will be infinite. During the training, the network searches for the minimum of the loss using the Adam optimizer \citep{Adam}, with an initial learning rate (the parameter that determines the size of the steps in the free parameter space) of 0.01. If the loss varies by less than $10^{-4}$ during 10 epochs, we divide the learning rate by 5 down to a minimum of $10^{-4}$, and the training ends when the loss has varied by less than $10^{-4}$ during 50 epochs.

The second step is the "validation" of the network, performed on a second \textit{validation set} of 0.5\% of the catalog ($5\times10^6$ pairs). During this phase the network defines a threshold $f_\mathrm{th}$: every pair resulting in $f>f_\mathrm{th}$ are considered as merging, while pairs with $f<f_\mathrm{th}$ are dubbed as chance superpositions. The value of $f_\mathrm{th}$ is defined by maximizing the $MCC$ on the validation set, consistently with the analysis discussed in the previous Sections.

Finally, once the network is trained and validated, we run it on a third \textit{test set} of 0.5\% of the catalog ($5\times10^6$ pairs\footnote{The network has not been run on the whole sample because of the computational cost of the test. Note however that the number of pairs used is large enough for this kind of architecture, \cite{Marchetti_17} typically had a sample with $2.5\times10^6$ objects.}). The test run performed results   in a very low $MCC$ ($\sim 0.1$), due to the extreme unbalance of the catalog (see \S\ref{sec:TwoExtremeCases}), that "teaches" the neural network to typically  answer that pairs never merge. 

To overcome this issue, we build a new \textit{balanced training set} ($2\times10^5$ pairs) containing 50\% of merging pairs and 50\% of non merging pairs. In order to check the good behavior of our network on a balanced catalog, we also build a \textit{balanced test set} ($3\times10^4$ pairs).
\begin{figure}
\begin{center}
\includegraphics[width=\columnwidth]{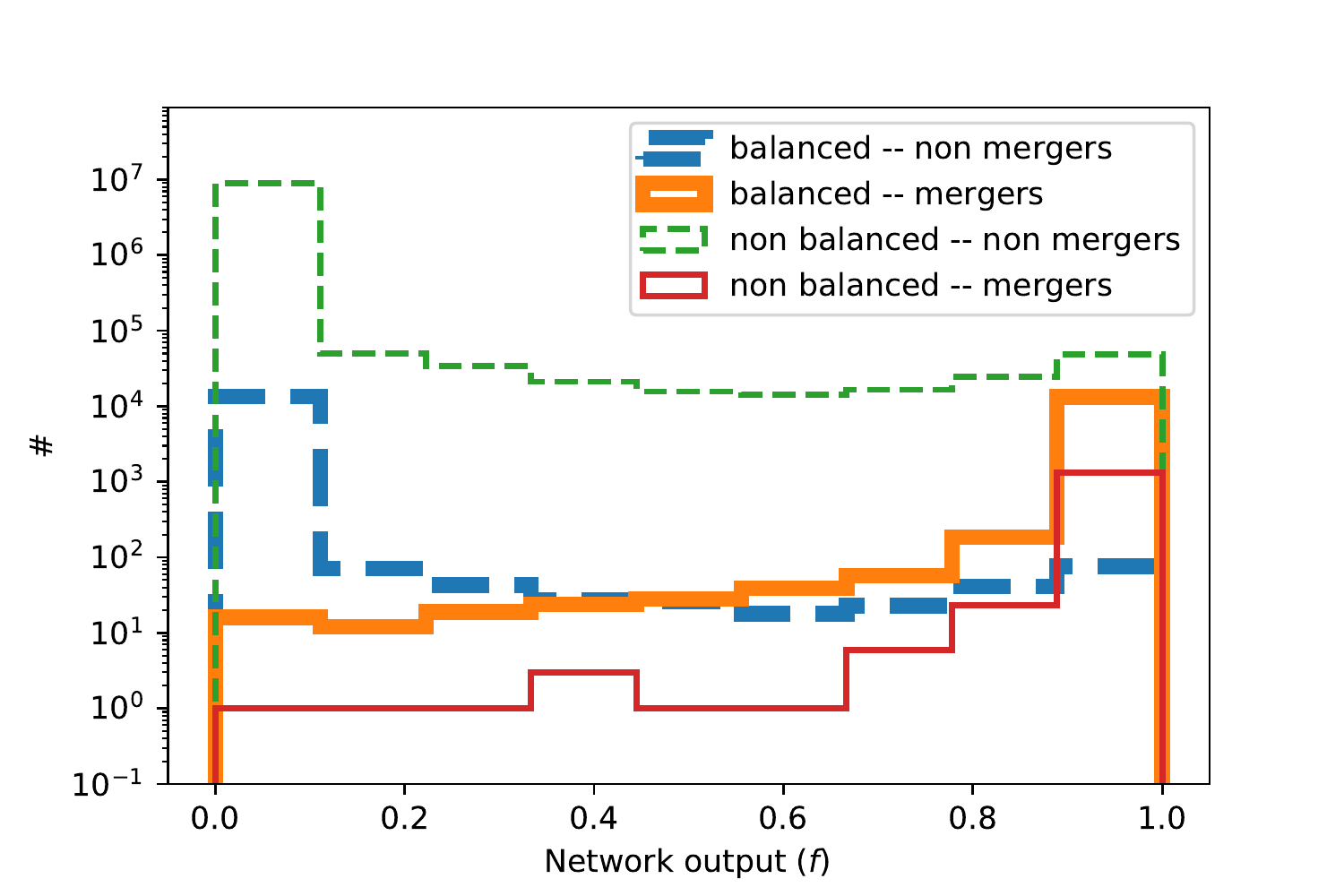}
\caption{Histogram of outputs from the network of mergers (solid lines) and non mergers (dashed lines) when applied to the balanced (thick line) and unbalanced (thin line) dataset. In most of cases, the network is able to classify correctly the pair, but a small fraction, which in is dominant (as the problem is unbalanced), of non merging pairs are identified as ``merging'' (as the problem is degenerate) which results in poor classification.}
\label{fig:histNN}
\end{center}
\end{figure} 

After training on the new balanced set and optimizing the threshold on the unbalanced validation set, we obtain a $MCC$ on the unbalanced test set of 0.41. This is again not much of an improvement compared with naive selection using only $\Delta z $ and $d$. The reason is that, among the large number of non merging pairs, a small fraction have similar properties than merging pairs (we recall that the problem is degenerate, see  \S\ref{sec:TwoExtremeCases}). However, as the problem is also unbalanced, although this fraction is small, the resulting number of false positive can be larger than the number of merging pairs itself. We show for instance in Fig.~\ref{fig:histNN} the histogram of outputs of the network, $f$, on mergers (solid lines) and non mergers (dashed lines) in the case of the balanced (thick line) and unbalanced (thin line) test sets. In both cases more than 90\% of mergers (non mergers) have $f>0.95$  ($f<0.05$): the network is perfectly capable of classifying most of pairs. In the case of the balanced test set, where degeneracy is minimized due to balancing, the network is excellent ($MCC=0.97$), however, in the case of the real dataset, the 0.35\% of non merging pairs with $f>0.95$, \ie classified as ``mergers'', outnumber the total number of pairs, resulting in a $MCC$ of 0.41. \hugo{This problem is inherent to the small number of input parameters in the network and, to obtain better results, more inputs should be used: one could think of parameters linked to morphology and/or disturbances, such as the $\phi$-asymmetries \citep{Conselice_00}, the Gini coefficient \citep{Lotz_08} or other reduced quantities to describe the image. However, given the large size of the dataset, using all the pixels of the images as input parameters in a more complex network would probably be the most efficient way of greatly improve the classification.  We postpone such analysis to a future study.}

\section{Conclusions}
\label{sec:conclusions}

In this paper, using the {\sc Horizon-AGN} simulation, we build a mock catalog of galaxy pairs in order to infer the optimal way to determine true merging pairs. We summarize our finding below:
\begin{itemize}
\item Using only the projected distance and redshift difference cannot be 100\% accurate: we found two pairs with similar projected properties but with different behaviors, some merging some not;
\item Nonetheless, the optimal parameters when using only the projected distance and redshift difference are $d_\mathrm{th}=86\kpc$ and $\Delta z_\mathrm{th} = 7\times 10^{-4}$. This result is in excellent agreemenet with the recent results of \cite{Ventou_19}. Note that the resulting $MCC$ (the metric we use in this paper) is only 0.38. This is due to a combination of both the degeneracy and unbalancing of the problem;
\item More detailed classifiers including the mass of the primary marginally improve the $MCC$ to 0.40. The improvement is much better for massive galaxies, because massive galaxies merge more frequently;
\item Including non-linear relations between the \hugo{9} parameters of each pair in the catalog (projected distance, redshift difference, masses, redshift, SFRs \hugo{and aspect ratios}) through a neural network again marginally improve the $MCC$ to 0.41. This confirms that the most relevant parameters to detect merging pairs are the projected distance and redshift difference. It also shows that, in order to be more predictive, future detection methods will need to use the full image instead of reduced quantities.
\end{itemize}

These new selection criteria can be used in large survey to refine the estimates of the evolution of the galaxy merger rate \citep[\eg][]{Ventou_19}, but also to study statistically the effects of mergers on the SFR \citep[\eg][]{Calabro_19} or AGNs \citep[\eg][]{Koss_12}.

\section*{Acknowledgments}
\hugo{HP is indebted to the Danish National Research Foundation (DNRF132) for support. HP, MV and YD acknowledge support from the European Research Council (Project no. 614199, `BLACK'). This work has made use of the Horizon Cluster hosted by the Institut d'Astrophysique de Paris; we thank Stephane Rouberol for running smoothly this cluster for us. HP thank A. Szenicer for insightful discussions. We thank the referee, Jon Loveday, for carefully reading the manuscript and providing useful comments.}

\appendix
\section{Changing our definition of merger}
\label{sec:ChangingOurDefitionOfMerger}

\begin{figure}
\begin{center}
\includegraphics[width=\columnwidth]{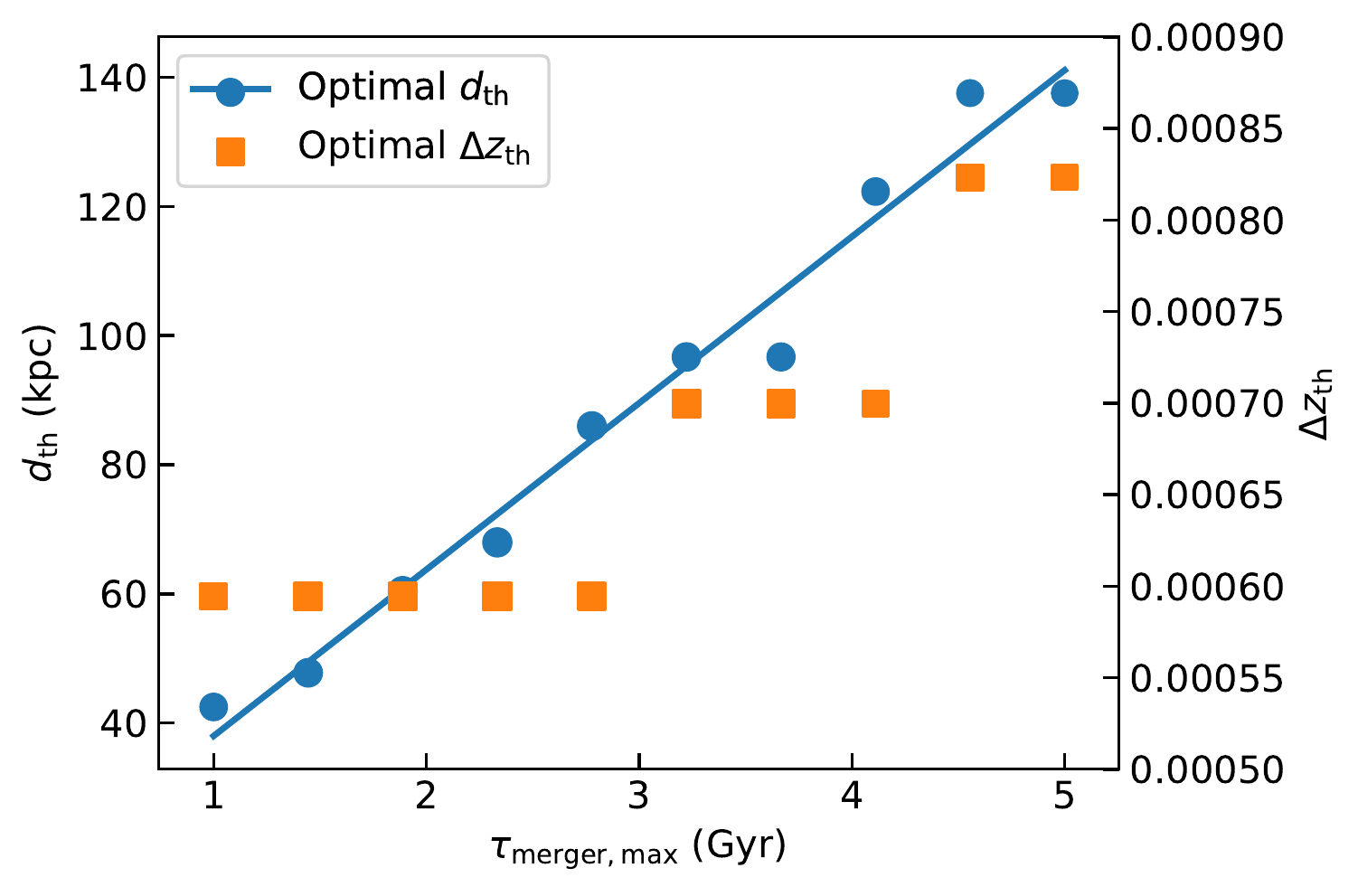}
\caption{Evolution of $d_\textrm{th}$ (blue dots) and $\Delta z_\textrm{th}$ (orange squares) with $\tau_{\rm merger, max}$, as well as the fit of Eq.~\eqref{eq:dOftau}.}
\label{fig:dthOftau}
\end{center}
\end{figure}

In \S\ref{sec:DetectionOfRealGalaxyMerger}, we specified that a pair was considered as ``merging'' if it merges within $\tau_{\rm merger, max}=3 \Gyr$ (or within $z=0$ if the cosmological time left is shorter than 3 Gyr). In this Appendix, we vary $\tau_{\rm merger, max}$ between 1 and 5 Gyr, and \hugo{see how this} affects our results. As we have found our three algorithms to have similar efficiency, we stick to the simplest one (\S\ref{sec:DependanceInDandZ}) and study how $d_\textrm{th}$ and $\Delta z_\textrm{th}$ vary with $\tau_{\rm merger, max}$. We show our results in Fig.~\ref{fig:dthOftau}. 

$\Delta z_\textrm{th}$ is not so affected by $\tau_{\rm merger, max}$, with a mean at $\sim 10^{-3}$, as found in \S\ref{sec:DependanceInDandZ}, and a standard deviation of 13\%. However, $d_\textrm{th}$ varies linearly as:
\begin{eqnarray}
\frac{ d_\textrm{th} }{ \kpc} = 26 \left( \frac{ \tau_{\rm merger, max} }{\Gyr}\right) + 12 \, . \label{eq:dOftau}
\end{eqnarray}
The $MCC$ is fairly constant, with 2\% variations and a mean at 0.38.

\section{Changing the depth of the lightcone}

\begin{figure}
\begin{center}
\includegraphics[width=\columnwidth]{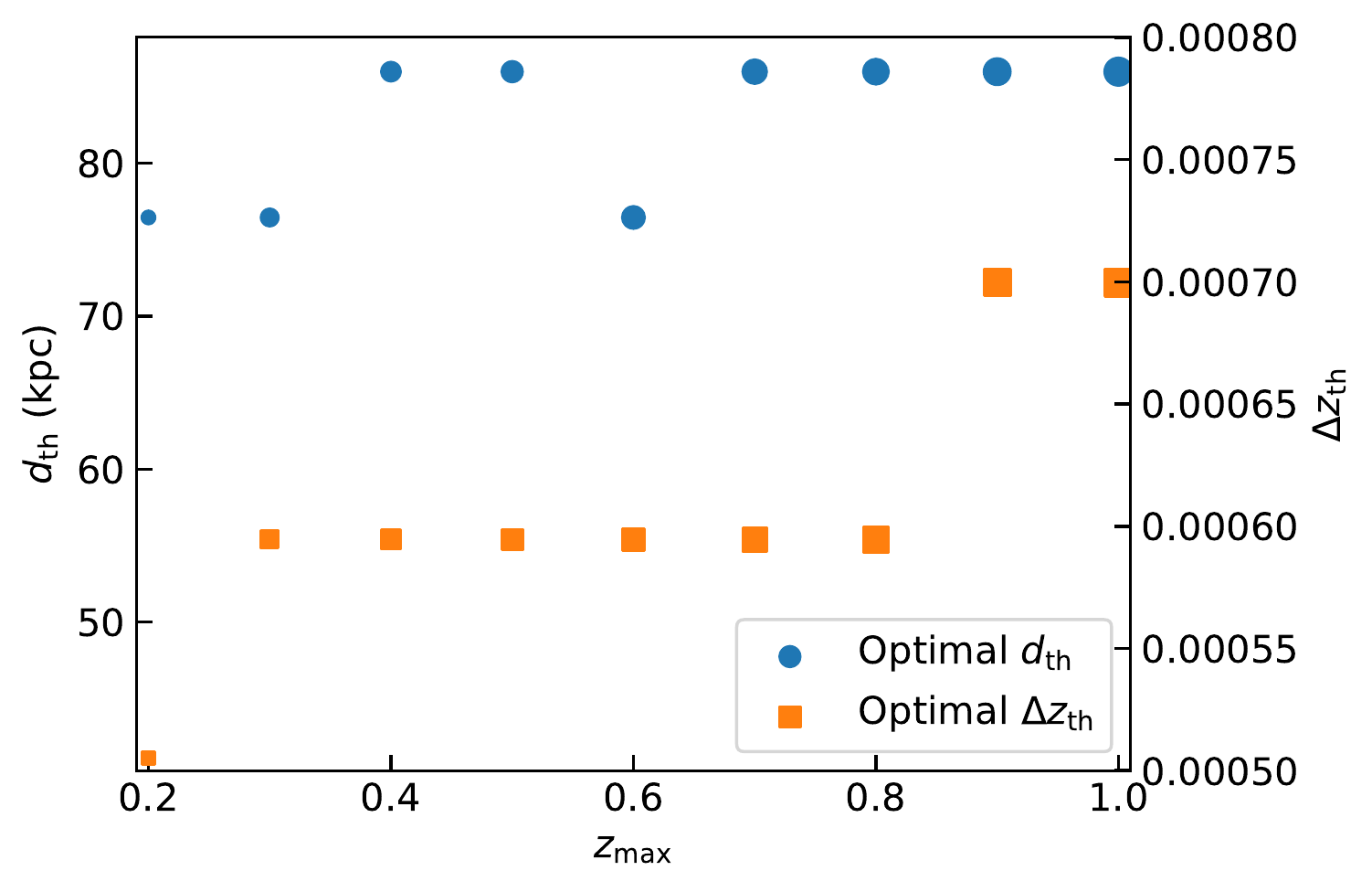}
\caption{Evolution of $d_\textrm{th}$ (blue dots) and $\Delta z_\textrm{th}$ (orange squares) with $z_\mathrm{max}$. The larger the dots the larger the $MCC$ (min: 0.25, max: 0.38).}
\label{fig:dthOfzmax}
\end{center}
\end{figure}

In \S\ref{sec:BuildTheCatalog}, we specified that we selected galaxies with $z<1$, however, for some reasons it is possible that real catalogs cannot achieve this maximal redshift. Vice versa, it is possible that real catalogs achieve higher redshifts. In both case, one can wonder if the thresholds have to be changed. In this Appendix, we select galaxies with $z_\mathrm{max}$ in [0.2, 1] and perform the same analysis as in \S\ref{sec:DependanceInDandZ}, assuming $\tau_{\rm merger, max}=3$. We show our results in Fig.~\ref{fig:dthOfzmax}.

Both for $d_\textrm{th}$ and $\Delta z_\textrm{th}$, little difference is found with variation of 18\%.

\bsp	\label{lastpage}
\end{document}